%% file: mblapp_v6sub.tex
\documentclass[aps,pra,twocolumn,showpacs]{revtex4-1}

\usepackage{amssymb,amsmath,amstext}                
\usepackage{graphicx}                                               
\usepackage{epstopdf}                                               
\usepackage{color}                                                     
\usepackage{bm}                                                        
\usepackage{appendix}                                              
\usepackage[utf8]{inputenc}
\usepackage{bbold}
\usepackage{bbm}
\usepackage{ulem}
\usepackage[dvipsnames]{xcolor}

\usepackage{latexsym}
\usepackage[colorlinks=true,citecolor=blue,linkcolor=magenta]{hyperref}

\usepackage{filemod}

\def\rmd{\mathrm{d}}
\def\be{\begin{equation}}
\def\ee{\end{equation}}
\def\bg{\begin{equation}\begin{gathered}}
\def\eg{\end{gathered}\end{equation}}
\def\B#1{\!\left(#1\right)}

\begin{document}

\title{Many-body localization for randomly  interacting bosons }

\author{Piotr Sierant} 
\affiliation{
Instytut Fizyki imienia Mariana Smoluchowskiego, 
Uniwersytet Jagiello\'nski, ulica Profesora Stanis\l{}awa \L{}ojasiewicza 11, PL-30-348 Krak\'ow, Poland}

\author{Dominique Delande}
\affiliation{Laboratoire Kastler Brossel, UPMC-Sorbonne Universit\'es, CNRS, ENS-PSL Research University, Coll\`ege de 
France, 4 Place Jussieu, 75005 Paris, France}
\affiliation{
Instytut Fizyki imienia Mariana Smoluchowskiego, 
Uniwersytet Jagiello\'nski, ulica Profesora Stanis\l{}awa \L{}ojasiewicza 11, PL-30-348 Krak\'ow, Poland}

\author{Jakub Zakrzewski}
\affiliation{Instytut Fizyki im. Mariana Smoluchowskiego, Uniwersytet Jagiello\'nski,  \L{}ojasiewicza 11, 30-348 Krak\'ow, Poland }
\affiliation{Mark Kac Complex
Systems Research Center, Uniwersytet Jagiello\'nski, Krak\'ow,
Poland. }
\email{jakub.zakrzewski@uj.edu.pl}

\date{\today}

\begin{abstract}
We study many-body localization in a one dimensional optical lattice filled with bosons. The interaction between bosons is assumed to be random, which can be realized for 
atoms close to a microchip exposed to a spatially fluctuating magnetic field. Close to a Feshbach resonance, such controlled fluctuations can be transfered to the interaction strength. 
We show that the system reveals an inverted mobility edge, with mobile particles at the lower edge of the spectrum. A statistical analysis of level spacings 
allows us to characterize the transition between localized and excited states. 
The existence of the mobility edge is confirmed in large systems, by time dependent numerical simulations using tDMRG. A simple analytical model predicts the long time behavior of the system. 
\end{abstract}

\maketitle
\section{Introduction}

Many-body localization (MBL) \cite{Basko06,Oganesyan07,Znidaric08} remains one of the challenging phenomena of many-body physics despite hundreds of papers per year appearing  on this subject (for recent reviews see \cite{Huse14,Rahul15}). One of the reasons is that MBL  breaks the common assumption  that many-body interacting systems should thermalize. 
For large isolated systems  the  eigenvector thermalization hypothesis \cite{Deutsch91,Srednicki94} suggests  that {\it local} observables thermalize in the following  sense: 
their averages do not contain information about the initial state after a sufficient thermalization time. This paradigm is not realized in many-body localized systems, 
where the local observables reveal a hidden memory in the system and remember their initial values. 

The ``standard model'' of MBL is the spin-1/2
 Heisenberg chain \be
\label{Hsta}
\hat{\mathcal{H}_{standard}}= J \sum_{i}^{L-1}\vec{S}_i\cdot\vec{S}_{i+1}
+ \sum_{i}h_i S_i^z
\ee
which, for random uniform $h_i\in[-H,H]$, shows a transition from an ergodic to MBL behavior for a sufficiently strong disorder ($H=H_c\approx3.5$ is an estimated transition 
disorder value \cite{Huse14}).
Using a Jordan-Wigner transformation, one can map the spin model  to a system of interacting fermions in a lattice, a favorable medium for cold atom experiments
that showed evidence for MBL in one-dimension (1D) \cite{Schreiber15} and two-dimensions (2D) \cite{Kondov15,Bordia16,Choi16}

Most MBL studies are based on exact diagonalizations \cite{Pal10,Luitz15,Luitz16} for small systems.
The basic understanding comes from the perturbative approach \cite{Basko06} based on Anderson localization of a single-particle model. 
Indeed, experimental results indicate that the localization border only weakly depends on the interaction strength \cite{Schreiber15}. The experiments up till now consider fermionic systems \cite{Schreiber15,Kondov15,Bordia16,Choi16,Lueschen17} - we shall consider bosons instead.

Is the single-particle localization a necessary ingredient? In a recent study considering bosons, we have shown \cite{Sierant17} that it is not the case. One may consider particles with 
random interactions. Such a system reveals MBL while, when the interactions are turned off, the randomness disappears and the system  has extended, single particle eigenstates.
In later works, a similar phenomenon was observed for fermions \cite{BarLev16,DasSarma17}. We shall consider the bosonic system in more detail here providing an understanding of the
observed MBL via a perturbative model, extending and clarifying the results reported in \cite{Sierant17}. Additional details will be presented elsewhere \cite{Sierant17b}.

\section{The model}

The Bose-Hubbard Hamiltonian describing a 1D system in an optical lattice within the 
tight binding approximation reads, assuming random on-site interactions \cite{Sierant17}
\bg
\label{H}
\hat{H}= -J \sum_{i}^{L-1}\B{\hat{a}_{i+1}^\dag \hat{a}_i + {\rm h. c.}}
+\frac{1}{2} \sum_{i}U_i\hat{n}_i\B{\hat{n}_i-1}, \\
[\hat{a}_i,\hat{a}_j^\dag]=\delta_{ij}, \ [\hat{a}_i,\hat{a}_j]=0, 
\  \hat n_i=\hat{a}_i^\dag\hat{a}_i,
\eg
with the first term describing the tunneling while the second term corresponds to interactions. 
Here, following \cite{Gimperlein05} we assume the interaction strength to depend on the site taking $U_i=Ux_i$ with $x_i$ being a random number uniformly distributed in $[0,1]$. We fix the energy (and time) scale by taking $J=1$. 

There are two standard approaches that help to identify the MBL phase. For relatively small systems (say at most 22 sites for the ``standard model'') one may apply exact diagonalization techniques to study properties of a given system in a considerable detail. Then both long-time dynamics as well as  properties of eigenstates and/or eigenvectors may be analyzed. Such studies necessarily suffer from finite size effects.  An alternative approach addresses the dynamics for large systems, similarly to experimental studies. Here, tDMRG techniques and its variants allow to simulate dynamics for quite large systems.
However, the time scale over which the dynamics may be followed reliably strongly depends on the properties of the system. In the MBL phase it has been shown on spin models that the entanglement entropy of an initial separable state grows at most logarithmically in time \cite{Znidaric08,Bardarson12}. That allows one to reach quite long times with standard algorithms.  The situation is more difficult 
in the critical region separating the MBL and the ergodic phase: here entanglement grows fast (power-like) \cite{Luitz16} limiting simulations to relatively short times. That makes predictions about the long time behavior of the system questionable.

With that in mind, we shall consider our model using both techniques: small size exact diagonalization as  well as tDMRG propagation for large system sizes. The complementary measures used in the two approaches shed some light on the localization phenomenon although our understanding of the MBL phase and especially of the  MBL-extended states transition is still far from complete. 

\begin{figure}
\includegraphics[width=1.0\linewidth]{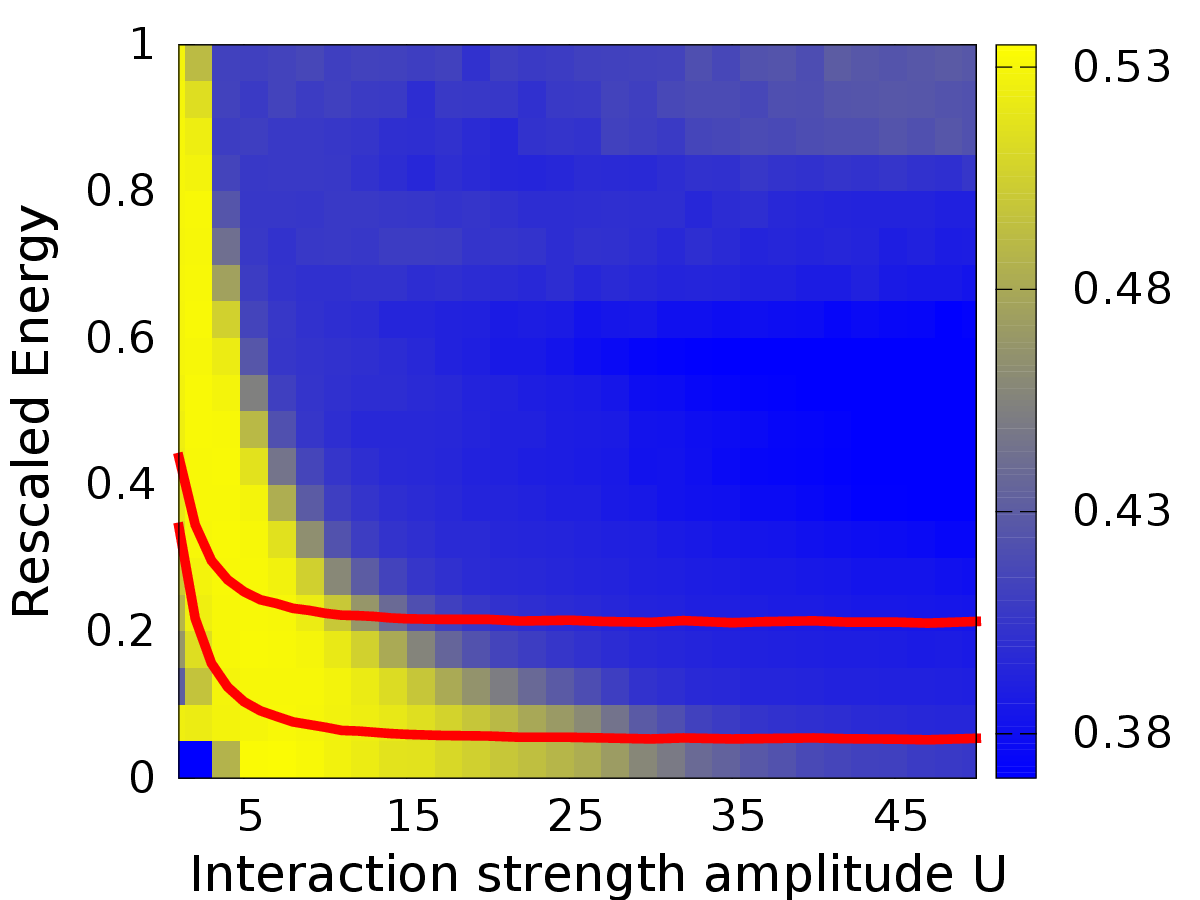}
\caption{Average ratio $\bar r$ between consecutive energy level spacings vs. the disorder amplitude $U$ and energy $\epsilon.$ High energy states (in blue) are close to the Poisson limit $\bar r=0.38$ indicating many-body localization, while low energy states (in yellow) are close to the GOE limit $\bar r \approx 0.53$ signaling extended states.
 Solid red lines correspond to the mean energy of the $|030303..\rangle$ (higher) and $|121212..\rangle$ (lower) states showing that, at intermediate $U$ values, the dynamics may be simultaneously ergodic (yellow) for the $|121212..\rangle$ state and localized (blue) for the $|030303..\rangle$ state. Data are collected for $L=6$ and $N=9$ with open boundary conditions for several realizations of the disorder. The blue square in the lower left corner is an artefact due to the very small number of levels in that region. 
} 
\label{fig:map}
\end{figure}

For our system, the occupation number of each single site can be up to the total number of particles, implying a large dimension of the local Hilbert space,
This compares unfavorably with the standard spin model \eqref{Hsta} (or spinless fermions) where the dimension of the local Hilbert space is fixed at 2.  In experiment with spinful fermions\cite{Schreiber15,Kondov15,Bordia16,Choi16,Lueschen17}, it is 4, still much less than for bosons. For that reason,  bosons are rarely discussed in the context of MBL, see however~\cite{Stolz16,Singh17}.

\begin{figure}
\includegraphics[width=1.0\linewidth]{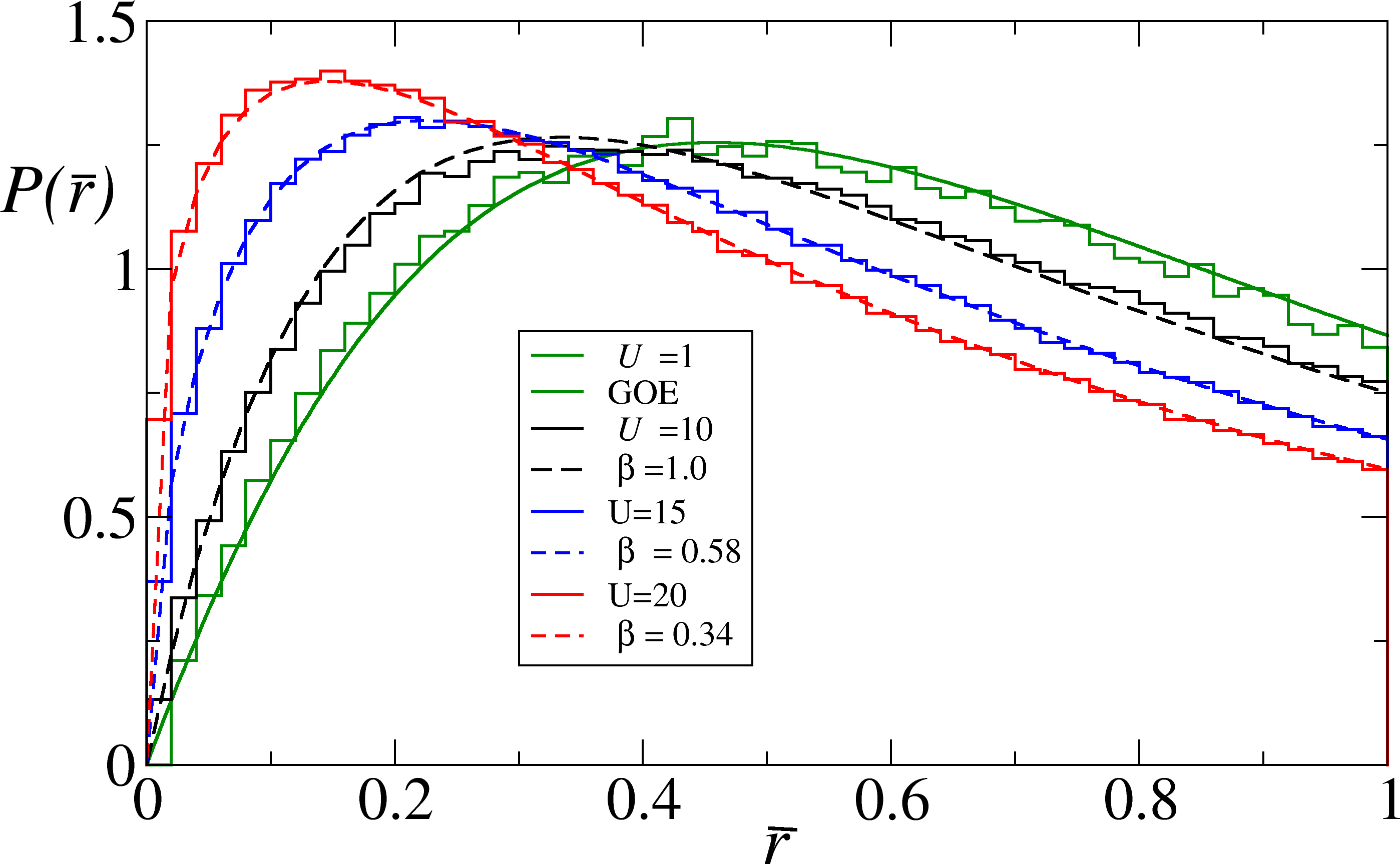}
\caption{Distribution of the ratio of consecutive energy level spacings, $P(r)$, for bosons with interactions randomly and uniformly distributed in the $[0,U]$ interval. Due to a significant dependence of system properties on energy (compare with Fig.~\ref{fig:map}) level spacings are collected in a narrow interval around the energy of the $|212121..\rangle$ state. The histograms present numerical results averaged over several realizations of the disorder. The green solid line is the prediction \eqref{rGOE} for the GOE ensemble of random matrices and reproduces the numerical results for $U=1.$ At higher $U$ values, the data are well fitted by 
the generalized semi-Poisson distribution \eqref{rsP} with the fitted repulsion parameter $\beta$ indicated in the Figure. All data are for $N=9$ particles on $L=6$ sites with open boundary conditions. }
\label{fig:rdis}
\end{figure}

\section{Small system sizes - level statistics approach}

\begin{figure}
\includegraphics[width=1.0\linewidth]{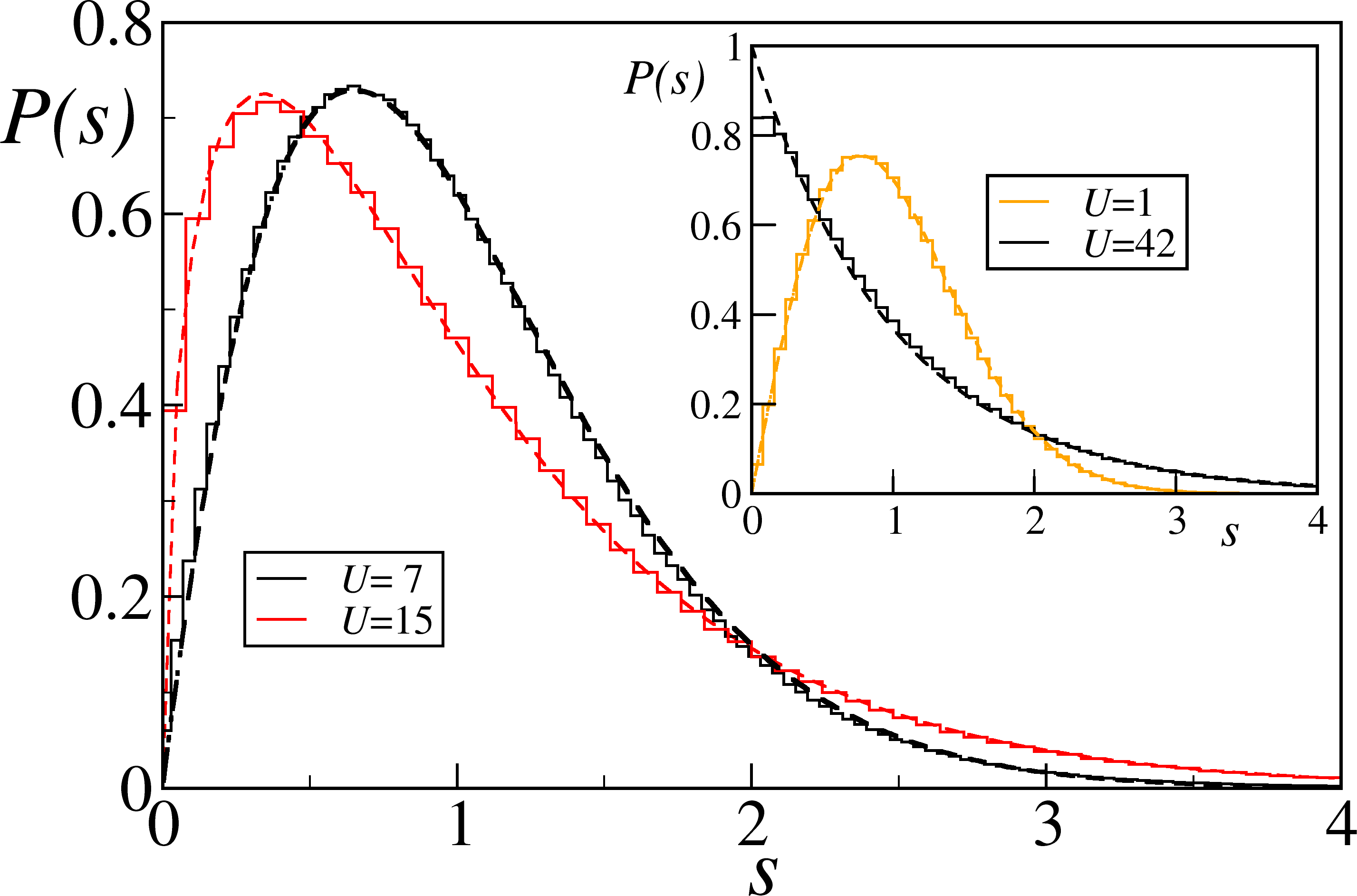}
\caption{Level spacing distributions for $N=9$ bosons on $L=6$ sites of a one-dimensional chain with open boundary conditions. Energy data in a narrow interval of energies around $|212121..\rangle$ state  are unfolded following the standard procedure. The $U=15$ data are well fitted by the generalized semi-Poisson distribution \eqref{gP} with {$\beta\approx0.6$}  while for {$U=7$} the data are well reproduced by a $P(s)\propto s\exp(-Cs^{2-\gamma})$ distribution, \eqref{serb}, proposed in \cite{Serbyn16}, with $\gamma\approx0.6$. The inset shows comparisons with limiting distributions: Poisson distribution for  $U=42$ and GOE distribution for $U=1$. } 
\label{fig:spa}
\end{figure}

The Bohigas-Giannoni-Schmidt conjecture \cite{bgs84} linked the regular or chaotic behavior of a classical system with statistical properties of the energy spectrum: ergodic systems are associated with the Gaussian Orthogonal Ensemble (GOE) (for time-reversal invariant systems) of random matrices. In the original formulation, the so called nearest-neighbor spacing statistics, $P(s)$, was considered that, for comparison with the GOE prediction, required unfolding of the spectrum, i.e. a smooth rescaling of the energy levels such that their mean spacing is equal to unity.  
Ten years ago, Oganesyan and Huse \cite{Oganesyan07} introduced a convenient dimensionless measure, the ratio of consecutive spacings $r_n=\delta_{n+1}/\delta_n$ with $\delta_n=E_n-E_{n-1}$ being the spacing between two consecutive 
energies. Actually it is common to consider the modified ratio $r\in(0,1)$,  defined as the minimum of $r_n$ and its inverse. Simplified closed formula
can be found for the distribution of $r$ in Gaussian Ensembles~\cite{Atas13} for small matrices, which are very close to the distribution for large matrices (the situation resembles here very much the famous Wigner distribution \cite{bgs84} for spacings, analytically available for $2\times2$ matrices, very close to the exact GOE distribution
for large matrices). For the GOE case, relevant for us, the approximate form is: 
\be
P(r) = \frac{27}{4}\frac{r+r^2}{(1+r+r^2)^{5/2}},
\label{rGOE}
\ee
yielding the mean $\bar r=0.53$.
At the other extreme, i.e. deeply in the localized regime, one expects uncorrelated energy levels with Poisson level spacing distribution. The corresponding $r$-distribution takes the form $P(r)=2/(r+1)^2$ with the mean $\bar r=2\ln 2 -1\approx0.386$ \cite{Atas13}. In the transition between localized and extended phases, one may be guided 
by propositions for the intermediate statistics \cite{Bogomolny99}. The semi-Poisson distribution (with linear level repulsion for small spacings and exponential behavior for large spacings) can be generalized to the case of
arbitrary repulsion $\beta\in[0,1]$ with $\beta=0$ corresponding to the Poisson distribution {and $\beta=1$ to the ordinary semi-Poisson distribution} \cite{Atas13b}. Those has been successfully used in the transition between MBL and extended   phase \cite{Serbyn16,Garcia-Garcia}. The corresponding
$P(r)$ may be analytically determined \cite{Atas13b} and is given by:

\begin{equation}
P(r)=\frac{2\Gamma(2\beta+2)\Gamma^2(\beta+2)}{(\beta+1)^2\Gamma^4(\beta+1)}\frac{ r^\beta}{( r +1)^{(2\beta+2)}}.
\label{rsP}\end{equation}

One must be, however, careful, when comparing numerical results with the theoretical distribution. The primary reason is that the system properties strongly depend on the energy. For a given disorder amplitude $U,$ we find all eigenvalues and rescale them to lie in $[0,1]$ interval. Then $\bar r(\epsilon)$ is found by averaging $r$ in a small energy window around the rescaled energy $\epsilon$ (with further averaging over many disorder realizations). This procedure, borrowed from \cite{Luitz15}, results in the color map plotted in Fig.~\ref{fig:map}. The blue color corresponds to $\bar r$ values close to the fully localized case $\bar r^{Poisson}=0.38$ while the yellow color corresponds to the ergodic phase with $\bar r^{GOE}=0.53$. Observe that, at low energy, the states have a tendency to delocalize, while higher energy states are generally more localized. Thus we reveal an unusual inverted mobility edge in the system. Its existence is quite easily understood: in our case, the disorder comes from random 
repulsive interactions, and strong 
disorder corresponds to high {interaction} energies.

Figure~\ref{fig:rdis} compares the numerically computed distribution of $r$ obtained for different disorder values with the fitted distributions of the form \eqref{rsP}. The energy data are taken in a narrow energy window around the energy corresponding to the initial state $|212121..\rangle$ used for temporal evolution (see below). This is quite important, as the system properties change with energy, see Fig.~\ref{fig:map}. The agreement of the numerical data with the generalized semi-Poisson distribution is excellent. For the  smallest $U$ (weak disorder and ergodic phase), the histogram coincides with the GOE prediction \eqref{rGOE}.

A similar comparison can be made for the level spacing distribution (after appropriate unfolding), $P(s)$, as shown in Fig.~\ref{fig:spa} -- see also \cite{Sierant17}. While the inset shows the limiting cases of GOE and Poisson distributions, 
the intermediate statistics in the transition regime is intricate. Close to the localized side ($U\ge10$), {one can use again the generalized semi-Poisson distribution (see above) whose prediction is}~\cite{Atas13b}
\be
P(s) = \frac{\Gamma(\beta+2)^{\beta+1}}{\Gamma(\beta+1)^{\beta+2}}s^\beta \exp\left[-\frac{\Gamma(\beta+2)}{\Gamma(\beta+1)}s\right]
\label{gP}
\ee
{smoothly evolving from a Poisson distribution ($\beta=0$) at $U=42$ to a semi-Poisson distribution ($\beta=1$) at $U=10$}. For smaller $U,$ we fit the distribution proposed by Serbyn and Moore \cite{Serbyn16} on the basis of a mapping to a plasma model:
\be
P(s)=C_1  s^\beta\exp(-Cs^{2-\gamma})
\label{serb}
\ee
where the two parameters $\beta$ and $\gamma$ are fitted while $C$ and $C_1$ are determined by the normalization and unit mean level spacing conditions. Observe that for $\gamma=1$ the Serbyn-Moore distribution 
\eqref{serb} reduces to a generalized semi-Poisson distribution \eqref{gP}. We have found that in the region of smaller $U<10$ close to the delocalized regime Serbyn-Moore ansatz with $\beta=1$ and fitted $\gamma$ reasonably well describes the numerical data (compare with Fig.~\ref{fig:spa}). 
Thus this distribution works well in the whole transition regime between MBL and ergodic phases. We have observed, however, that the regions of significant changes of $\beta$ and $\gamma$ are quite distinct.
On the localized side, $\gamma=1$ and the spacings reveal an exponential tail for large $s$. In that region $\beta$ changes smoothly from a full  
Poisson (MBL) limit with $\beta=0$ to the semi-Poisson limit with $\beta=1$.
Going further into the delocalized regime (smaller $U$ in our case) $\beta=1$ but $\gamma$ decreases to 0 reaching a GOE Gaussian tail in the fully ergodic regime. In the transition region, for $U\in [10,17],$ slightly better fits are obtained fitting simultaneously $\beta$ and $\gamma$. Bearing in mind that \eqref{serb} is necessarily an approximate fitting formula -- reducing e.g. for $\beta=1$, $\gamma=0$ to the $2\times 2$ Matrix approximate Wigner distribution \cite{bgs84} -- we present one parameter fits only as they work quite well.

Let us summarize the results obtained from the statistical analysis of levels for systems of small size. The system of bosons with random interactions reveals a pronounced inverted mobility edge: states with lower energy localize at larger disorder strength. In the transition region between the ergodic and localized phases, the Serbyn and Moore spacing distribution \eqref{serb} reproduces our numerical results. In particular two transition regions have been identified: the ``more localized'' region with a generalized semi-Poisson statistics (varying $\beta$, $\gamma=1$ in \eqref{serb}) and a region touching the ergodic part (with $\beta=1$  and varying $\gamma$). The same distribution works in the transition regime for the spin model considered in \cite{Serbyn16} and for our diagonal but nonlinear (as appearing in the interaction term) disorder for bosonic system.

\section{Time-dependent dynamics and the persistence of nonergodic character}

In experimental studies of MBL, a reliable access to level statistics is a formidable task and has not been attempted up till now. Instead, the experiments concentrate on the nonergodic behavior of local observables. Their average values at long times provide an evidence that the system remembers its initial state. This approach has been initiated 
in the Munich experiments \cite{Schreiber15} where in the initial state every second site of the optical lattice was prepared void of fermions. Thus fermions fill e.g. even sites while odd sites remain empty. The system then evolved in the presence of disorder. In the ergodic situation, one expects that the population of odd $N_o=\sum_{i} n_{2i+1}$
 and 
even sites $N_e=\sum_{i} n_{2i}$ equalize. We define the imbalance ${I}(t)$ as

\be
{I}(t)=\frac{N_e(t)-N_o(t)}{N_e(t)+ N_o(t)}.
\label{imb}
\ee
The experiment \cite{Schreiber15} has revealed that indeed, for a sufficiently strong disorder, the imbalance does not decay to zero at long time.

We follow the path indicated by experimentalists and calculate the imbalance for our bosonic system. We take as initial state the product of Fock states on each lattice site $|\Psi\rangle=|n_1n_2n_3..\rangle$ with $n_i$ being the occupation of site $i$.
In particular, we use the state $|\Psi_1\rangle = |121212..\rangle$ (we multiply the imbalance \eqref{imb} by 3 to have it equal to unity at $t=0$). The time evolution is carried out using a home-made tDMRG code \cite{Schollwoeck11,Vidal03,Vidal04,Zakrzewski09} which allows us to treat systems of reasonable size. We report here the data for $N=90$ bosons on $L=60$ sites. The detailed time dependence
was presented in~\cite{Sierant17}. The typical $I(t)$ contains an initial transient after which, deeply in the MBL regime, it stabilizes at a finite value (depending on the disorder strength), with small
short time fluctuations as well as a significant dependence on the disorder realization. To smooth out these fluctuations, we average the final result over 20 disorder realizations and over time.
Typical runs reach times $tJ=50$ and the data are averaged over the $tJ\in[30,50]$ interval. The reader is advised to consult \cite{Sierant17,Sierant17b} for details of the time dependence, as well as for the evidence that
the entanglement entropy grows logarithmically in time, which is one of the smoking guns for MBL \cite{Znidaric08,Bardarson12}. Here we concentrate on the dependence of the long time imbalance vs. disorder.

\begin{figure}
 \includegraphics[width=1.0\linewidth]{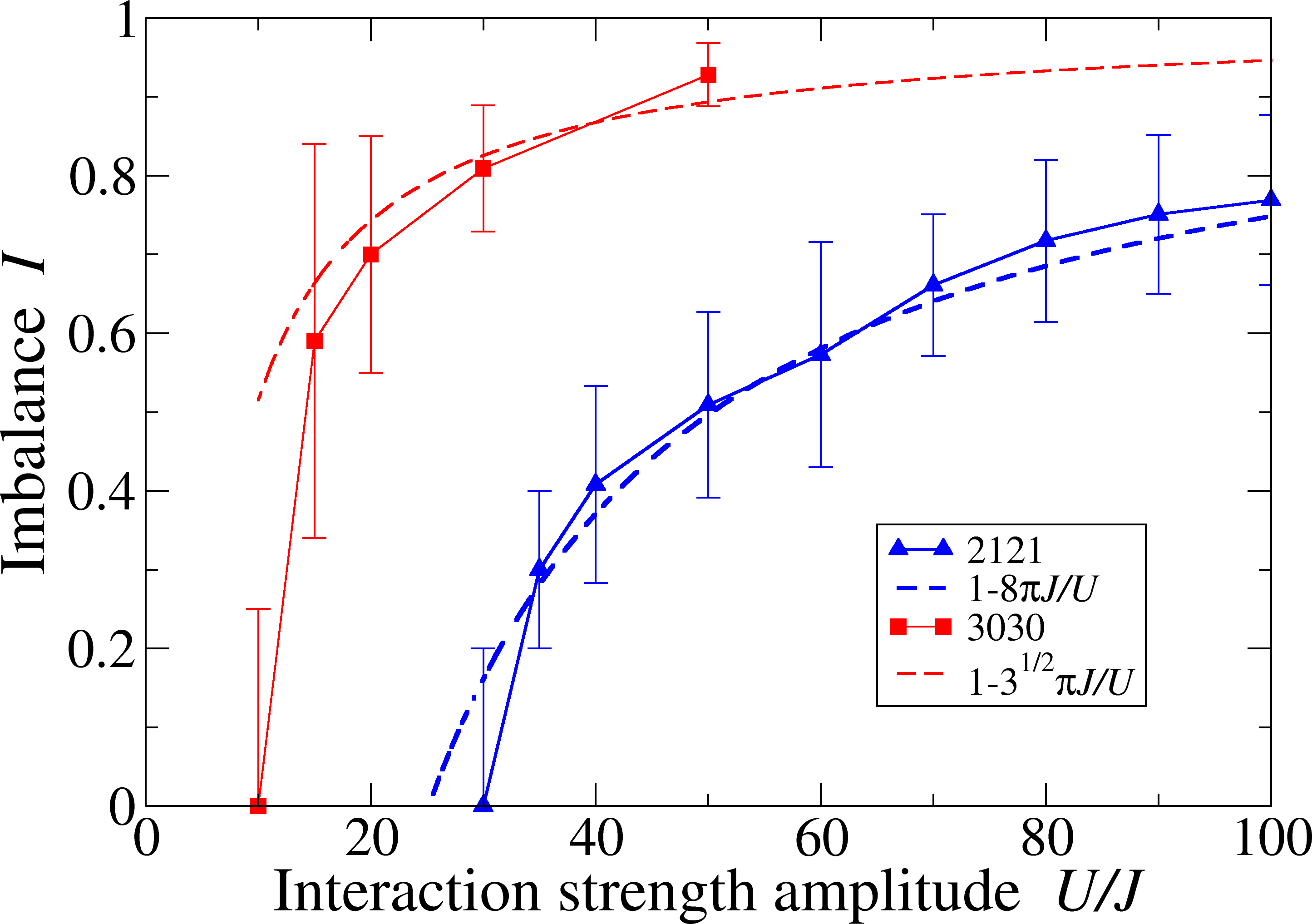}
\caption{
	Long time imbalance (\ref{imb}) vs. disorder strength for two different initial states
	$|1212..\rangle$ and $|0303..\rangle$, obtained by numerical propagation over time using the tDMRG algorithm. A non-zero value at large $U$ indicates many-body localization. The difference between the two curves shows that the localization properties strongly depend on energy, in agreement with Fig.~\ref{fig:map}.
	The dashed lines are the analytic predictions of a simple two-site model,  Eqs.~\eqref{Eq:mag21} and \eqref{Eq:mag30}.
} 
\label{fig:evo}
\end{figure}

Figure~\ref{fig:evo} presents the imbalance as a function of the disorder strength for the $|\Psi_1\rangle=|121212..\rangle$ and $|\Psi_2\rangle\equiv|030303..\rangle$ initial states.
Those state lie in different energy range. Observe that the energy of $|\Psi_1\rangle$ is $E_1=\sum_{i} U_{2i}$ while that of  $|\Psi_2\rangle$ equals $E_2=\sum_{i} 3U_{2i}$. We observe that the imbalance $I$ depends strongly on the energy. In particular, for $U\in[10,30]$,  $|\Psi_2\rangle$ shows a significant long-time imbalance indicating MBL while for $|\Psi_1\rangle$, the imbalance vanishes.
The large error bars indicate fluctuations over individual disorder realizations. For parameters leading to low imbalance values, the spreading of entanglement limits
the final time to $tJ=10-15$ and the tDMRG runs use a lot of CPU time and computer memory.

The dashed lines are the analytic predictions obtained using a simplified two-level scheme. described in detail in the next Section.

\section{Final imbalance: two-site approximation at large {\it U}}

The idea behind this approach is that, for very large $U,$ most sites are isolated from the neighbors because the on-site energies
differ my much more than $J,$ inhibiting any hopping and transport. The only contribution to transport are the rare cases where neighboring sites have almost the same on-site energies. The approximation thus consists in restricting the dynamics to pairs of 
quasi-resonant neighboring sites. This two-level approximation gives rise to Rabi oscillations between the two sites. On the average, it transfers some population from the initially populated state to its neighbor.

Consider first the approximate dynamics for the $|121212..\rangle$ initial state. For large $U,$ one may expect that it couples predominantly in the Hilbert subspace spanned by states
with also unit or double occupation but realized in different order. 
The 2x2 matrix representing the Hamiltonian on two consecutive sites in the $|21\rangle,|12\rangle$ basis writes:
\be
\begin{pmatrix}
 U_1 & -2J \\
 -2J & U_2
\end{pmatrix}
\ee
It is straightforward to show that, during the temporal evolution from the initial state $|12\rangle$, the average
populations in the two sites are:
\begin{eqnarray}
 n_1 & = & \frac{8J^2}{(U_1-U_2)^2+16J^2}+1\nonumber \\
 n_2 & = & \frac{(U_1-U_2)^2+8J^2}{(U_1-U_2)^2+16J^2}+1
\end{eqnarray}
leading to an average imbalance:
\begin{equation}
 n_2-n_1 = \frac{(U_1-U_2)^2}{(U_1-U_2)^2+16J^2}
\end{equation}
which is obviously zero in the resonant case $U_1=U_2$ and unity in the far off-resonant case $J\ll |U_1-U_2|.$

What is needed is to average this imbalance over the distributions of $U_1$ and $U_2,$ that is:
\begin{equation}
 \overline{n_2-n_1} = \frac{1}{U^2} \int_0^{U} \int_0^{U}\ \frac{(U_1-U_2)^2}{(U_1-U_2)^2+16J^2}\ \rmd U_1\ \rmd U_2 
\end{equation}
The integral can be computed going to the sum/difference variables $U=U_1+U_2,X=U_1-U_2:$
\begin{equation}
 \overline{n_2-n_1} = \frac{1}{2U^2} \int_{-U}^{U}{\ \rmd X\ \frac{X^2}{X^2+16J^2}\ \int_{|X|}^{2U-|X|}{\ \rmd U'}} 
\end{equation}
leading to 
\begin{equation}
 \overline{n_2-n_1} = 1\ -\ \frac{8J}{U} \arctan \frac{U}{4J}\ +\ \frac{16J^2}{U^2} \log\left(1+\frac{U^2}{16J^2}\right)
\end{equation}
In the limit $U\ll J,$ it correctly gives $\overline{n_2-n_1}=0.$ More interestingly, in the limit $U\gg J,$ it gives
$\overline{n_2-n_1} = 1-4\pi J/U.$

Finally, it must be taken into account that the initial population on the even site can be transfered to either the neighboring left or right site. As the two processes are essentially independent, this doubles the population depletion, finally leading to the prediction for the imbalance:
\begin{equation}
 {I} = 1\ -\ \frac{8\pi J}{U} 
 \label{Eq:mag21}
\end{equation}

Similar arguments may be used for the $|030303..\rangle$ initial state. The main coupling is to transfer one boson of an occupied site (leaving 2 bosons on the site) to the neighboring site.  In a crude approximation neglecting further couplings outside the two state subspace, the matrix in the $|12\rangle,$ $|03\rangle$ basis reads:
\be
\begin{pmatrix}
 U_1 & -\sqrt{3}J \\
 -\sqrt{3}J & 3U_1
\end{pmatrix}
\ee

Following exactly the same reasoning as above (a single integral over disorder is needed only) and, as before, taking into account that triple occupation may decay to both sides, yields the prediction for the final imbalance as 

\begin{equation}
{I} = 1-\frac{\sqrt{3}\pi J}{U} 
 \label{Eq:mag30}
\end{equation}

Both predictions \eqref{Eq:mag21} and \eqref{Eq:mag30} are compared with numerical
results in Fig.~\ref{fig:evo}. They work surprisingly well indicating that the observed localization is quite strong and the corresponding localization length cannot exceed 1-2 sites. 

\section{Conclusions}

We have shown that it is possible to observe MBL for interacting bosons with random interaction strength. In such a system the disorder comes from interactions only. In other words, without disorder, the system possesses extended states only. That suggests that the observed MBL is of non-perturbative character. Still, very simple models based on two-site approximations yield accurate predictions for the long-time imbalance, indicating that, at least for strong disorder (strong interactions), the MBL length in space does not exceed few sites.
Numerical data {from exact diagonalization} for small systems (where statistical properties of eigenvalues were considered) and from temporal evolution for large systems 
indicate that the system possesses an unusual inverted  mobility edge. 
A comparison with the more standard disorder with random chemical potential will be presented elsewhere \cite{Sierant17b}.


\acknowledgments
We enjoyed discussions with Fabien Alet and Antonello Scardicchio.
This work was performed with the support of EU via Horizon2020 FET project QUIC (nr. 641122). Numerical results were obtained with the help of PL-Grid Infrastructure. We acknowledge support of the National Science Centre, Poland via project No.2015/19/B/ST2/01028 (PS and JZ).

\input{mblapp_v5.bbl}
\end{document}

%% file: mblapp_v5.bbl
%